\documentclass[amsmath,amssymb,twocolumn,aps,pra,floatfix]{revtex4}
\usepackage{graphicx,amsmath,epsfig,epstopdf}% Include figure files
\usepackage{dcolumn}% Align table columns on decimal point
\usepackage{bm}% bold math
\usepackage{lipsum}
\newcommand{\bra}[1]{\ensuremath{\left< #1 \right|}}
\newcommand{\ket}[1]{\ensuremath{\left| #1 \right>}}

\newcommand{\asd}{\begin{equation}}
\newcommand{\dsa}{\end{equation}}
\newcommand{\stwo}{\text{\it Strategy 2}\,}
\newcommand{\sone}{\text{\it Strategy 1}\,}
\usepackage{mathrsfs}
\usepackage{palatino}
\begin{document}

\title{Non-Markovianity and System-Environment Correlations in a Microscopic Collision Model}%
%\thanks{A footnote to the article title}%

\author{Ruari McCloskey and Mauro Paternostro}
\affiliation{Centre for Theoretical Atomic, Molecular and Optical Physics, School of Mathematics and Physics, Queen's University, Belfast BT7 1NN, United Kingdom}

\date{\today}% It is always \today, today,
             %  but any date may be explicitly specified

\begin{abstract}
We show that the use of a recently proposed iterative collision model with inter-environment swaps displays a signature of strongly non-Markovian dynamics that is highly dependent on the establishment of system-environment correlations. Two models are investigated; one in which such correlations are cancelled iteratively and one in which they are kept all across the dynamics. The degree of non-Markovianity, quantified using a measure based on the trace distance, is found to be much greater for all coupling strengths, when system-environment correlations are maintained. %The responsibility for SECs in the non-Markovianity of the system is also backed up with observations about the contributions to the non-Markovianity investigated with the upper bound to the trace distance derivative. %Also, the model is investigated using Monte Carlo techniques and the measure of non-Markovianity is found to be fairly stable at high probabilities of system-ancilla interaction and high inter-ancilla swap strength.
\end{abstract}
% PACS, the Physics and Astronomy
                             % Classification Scheme.
%\keywords{Suggested keywords}%Use showkeys class option if keyword
                              %display desired
\maketitle

One of the obstacles to the grounding of fully scalable quantum technologies is, notoriously,  the subversion of {decoherence}, %For some, decoherence in the macroscopic limit is interesting because it describes the classical world in which we all live. However, from a quantum information viewpoint, 
that is, the process of losing the information encoded in a quantum-mechanical system due to its interaction with an environment~\cite{Palma96,Breuer02}. The uncontrollable process of exchange of information between a quantum system and its environment is responsible for the consequent degradation of the (quantum) coherence in the state of the former. %In an attempt to somehow arrest this process, one first thinks to attempt to isolate the quantum system, but it is impossible to fully isolate a quantum system from its environment and therefore impossible to stop decoherence from occurring.
The study of quantum decoherence has been the focus of a considerable body of investigations, both at the theoretical and experimental level, aimed at understanding and ultimately taming the effects that system-environment interactions have on a given dynamics. 

More recently, inspired also by the improved control over small-scale solid-state quantum devices, whose open-system dynamics is made difficult by the non-flat inherent {\it structure} of the environment that surrounds them, considerable attention has been given to the characterisation of the fundamental differences between Markovian and non-Markovian open-system dynamics~\cite{Breuer02,Breuer12}, a task that has been empowered by the formulation of useful tools for the quantitative assessment of the features of non-Markovianity~\cite{Breuer09,varia} which have enabled related investigations in various directions~\cite{varia2}, including test-bed simulations of explicitly non-Markovian evolutions and the experimental study on the transition from one dynamical regime to the other~\cite{simulations}. 

% Instead of trying to subjugate the flow of information to the environment, in recent years, a lot of work has been done in trying to create a system in which there is observable {\it feedback of information} from the environment to the quantum system~\cite{Palma12}. This information feedback is known as {\it non-Markovianity} due to the dynamical differences in these models from fully Markovian (i.e stochastic) processes, which exhibit no memory retention~\cite{Vacchini12}. In these cases the dynamics is governed by a Lindblad-type master equation ~\cite{Breuer02,Dumcke}. Markovian dynamics are, in general, only an approximation to the actual dynamics~\cite{daniel94}, which should be non-Markovian and indeed there are several cases known in which the Markovian approach falls apart completely~\cite{Michel05}. 
The idea behind most of the proposed measures of non-Markovianity is the quantification of a ``backflow" of some type. If the environment is able to retain and feed-back to the system it is interacting with part (or all) of the ``information'' (intended in a very broad sense) that the latter has previously poured into the former during the evolution, quantitative figures of merit can be identified that are able to signal the resulting dynamics as non-Markovian~\cite{Breuer09,varia}. One such indicator~\cite{Breuer09} is based on the use of the trace distance $D\left[\rho_1,\rho_2\right]$~\cite{Nielsen,Bengtsson}, which quantifies the degree of distinguishability between two arbitrary quantum states $\rho_1$ and $\rho_2$. When subjected to the same physical open-system model, two different states would in general undergo different evolutions. Trace distance is contractive under positive trace preserving maps, which is a property that enters significantly into the definition of non-Markovian quantum dynamics put forward in Ref.~\cite{Breuer09}: any increase in the trace distance should be taken as a fingerprint of non-Markovianity. %The total increase in the trace distance over the dynamics is the Breuer measure for non-Markovianity~\cite{Breuer09,Laine10}. Despite these advances in definitions of quantum non-Markovianity, 
Although a unified view on the reasons behind the emergence of non-Markovian features has not yet been found, progress has been made in the establishment of a hierarchical relation among some of the measures proposed so far~\cite{tony,cruscinski} and a careful assessment of the trace distance-based measure has allowed us to pinpoint the emergence of system-environment correlations (SECs)  established across the dynamics as being key for non-Markovian dynamics.

A striking case in which such backflow can be studied in great detail is the class of collision-based mechanisms for the microscopic modelling of system-environment interaction (including correlated baths), from Markovian evolution~\cite{Buzek01, Buzekothers,Palma12} all the way to explicitly non-Markovian ones~\cite{Ciccarello13, Ciccarello}. In such models, a quantum system $S$ {\it collides} sequentially with the elements of a multi-article environment, whose constituents might or might not mutually interact. Each collision results in an inevitable pouring of information from the system into an environmental element. The resulting system's dynamics is fundamentally dependent upon whether or not such intra-environment interactions take place~\cite{Buzek01, Buzekothers,Palma12, Ciccarello13,Ciccarello}. Suitable conditions (on the form and nature of the system-environment and environment-environment interaction and preparation) can be established that favor the emergence of strong non-Markovian features.

In this paper, we dig into the analysis of the collisional model by studying the role played, in the establishment of overall non-Markovian features, by the correlations that are set between the system and the environmental element with which it has interacted. In our analysis, we take the viewpoint provided by the trace norm-based measure of non-Markovianity~\cite{Breuer09}, thus putting us in a perfect position to address the overall role played by SECs in the generation of non-Markovianity. By tracking the evolution of SECs as the system interacts with the elements of a multipartite environment, we are able to pinpoint a fundamental difference in the way system-environment interaction is modelled, which in turn results in quantitatively different manifestations of non-Markovianity. Our analysis is confirmed by a numerical ``experiment", implemented using quantum Monte Carlo methods, that renders the system-environment collision process intrinsically aleatory, and thus closer to the actual way this interaction would take place in a physical system. This study paves the way for interesting investigations on the way memory effects retained in a system-environment interaction mechanism affect, say, figures of merit of thermodynamical relevance, such as heat exchanged with the environment and work performed on or by the system. 

\section{Description of the collision model}
\label{model}

We consider the dynamics of a qubit system (labelled $S$) with logical states $\{\ket{0},\ket{1}\}_S$ that interacts sequentially with the elements $\{E_1,E_2,..,E_N\}$ of an N-party environment (which we dub {\it super-environment} hereafter). The label that identifies each sub environment is provided by the position occupied by the corresponding environmental element in a spatial lattice. While the elements of the super-environment could obviously be of any nature, for the sake of illustration here we focus on the case of simple two-level systems whose logical states are labelled as $\{\ket{0},\ket{1}\}_{j}$~($j=1,..,N$). 

Needless to say, as far as the interaction between $S$ and the $j^\text{th}$ element of the super-environment is concerned, many choices are available. Here we concentrate on the case of a coherent interaction (i.e. a mechanism that can be described by a Hamiltonian model of some form) giving rise to the unitary time-evolution operator
\begin{equation}
\hat{\cal U}_{S,j}(\gamma)=(\cos\gamma)\hat\openone_{S,j}+ i(\sin\gamma)\hat {\cal S}_{S,j}.
\end{equation}
Here $\hat\openone_{S,j}$ is the identity operator,  $\gamma\in{\mathbb R}$, is a dimensionless interaction strength for the interaction, and we have introduced the swap gate $\hat {\cal S}_{S,j}$ that, in the ordered basis $\{\ket{k,l}_{S,j}\}$ (with $k,l=0,1$) reads~\cite{Nielsen} 
\begin{equation}
\hat{\cal S}_{S,j}=
\begin{pmatrix}
1&0&0&0\\
0&0&1&0\\
0&1&0&0\\
0&0&0&1
\end{pmatrix}.
\end{equation}
A similar model rules the interaction between two nearest-neighbor elements of the super-environment. That is, contrary to the case addressed in Ref.~\cite{Ciccarello13} in which a stochastic process regulates the subenvironment-subenvironment coupling, here we consider the unitary evolution 
\begin{equation}
\hat{\cal E}_{j,j+ 1}(\delta)=(\cos\delta)\hat\openone_{j,j+1}+ i(\sin\delta)\hat{\cal S}_{j,j+1}
\end{equation}
with $\delta\neq\gamma$, in general, and $\hat\openone_{j,j+1},\hat{\cal S}_{j,j+1}$ the analogue of the operations introduced above, defined in the Hilbert space of the elements $j$ and $j+1$ of the super-environment. Such interactions give rise to the dynamical maps
\begin{equation}
\begin{aligned}
&\hat\Phi_{S,j}[\rho]=\hat{\cal U}_{S,j}(\gamma)\,\rho\,\hat{\cal U}^\dag_{S,j}(\gamma),\\
&\hat\Psi_{j,j+1}[\rho]=\hat{\cal E}_{j,j+1}(\delta)\,\rho\,\hat{\cal E}^\dag_{j,j+1}(\delta),\\
\end{aligned}
\end{equation}

%%%%%%%%%%%%%%%%%%%%%%%%%%%%

\begin{figure}[t]
\includegraphics[width=\columnwidth]{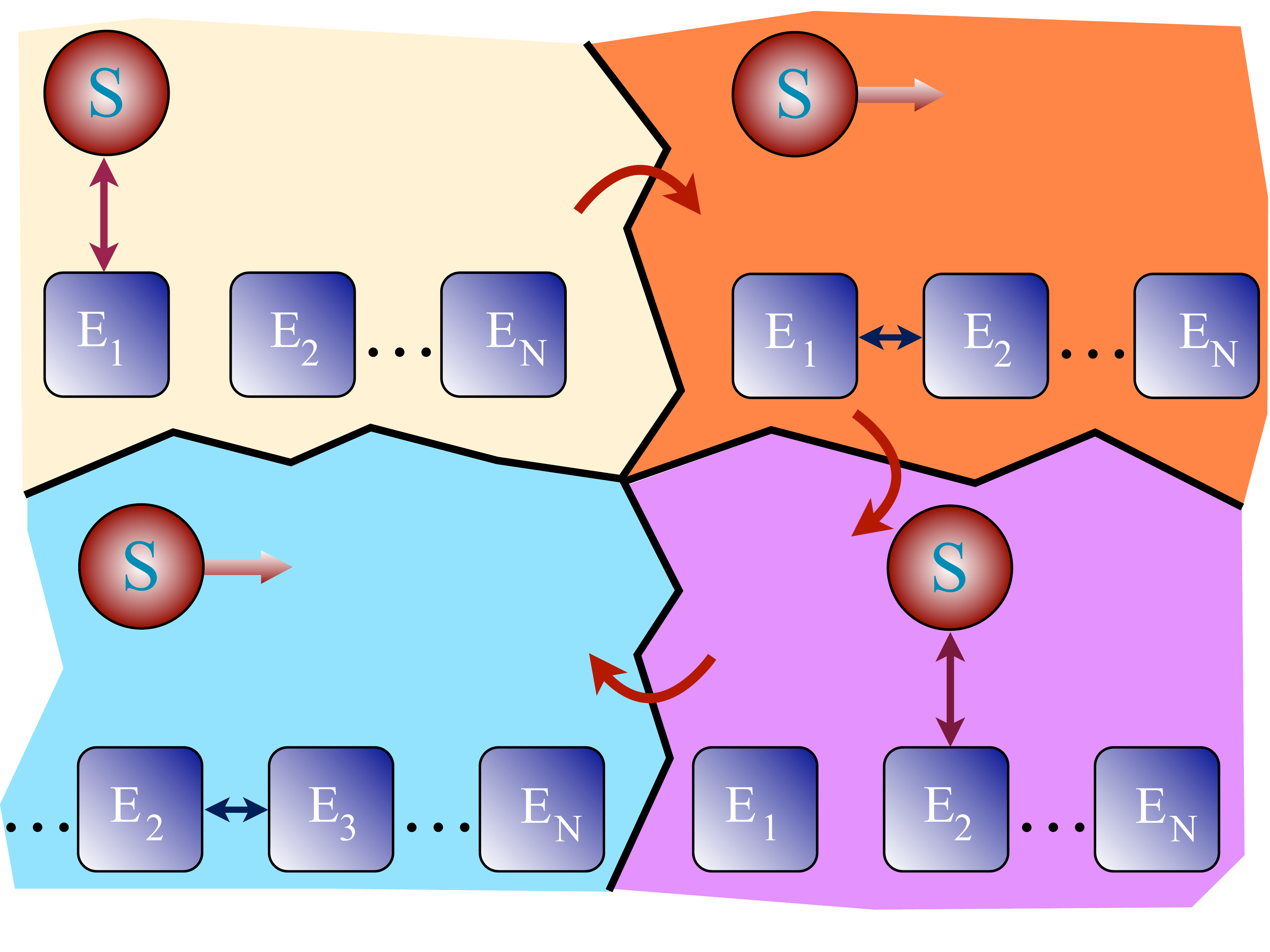}
\caption{Pictorial sketch of the iterative process at the basis of the collision model for system-environment interaction. System $S$ interacts sequentially with the elements of an $N$-party environment consisting of subsystems $\{E_j\}~(j=1,..,N)$. Following the sense of the inter-panel arrows, we distinguish four stages: $1)$ $S$ interacts with the first element $E_1$ of the environment. It then shift by one site, while $E_1$ interacts with its rightmost nearest neighbour $E_2$ [step $2)$]. In step $3)$, $S$ interacts with $E_2$ and then moves forward [step $4)$], while $E_2$ and $E_3$ interact. The specific form of the $S-E_j$ and $E_j-E_{j+1}$ interaction considered in our analysis is given in the body of the manuscript.  
%The basic collision model (left) and the model with inter-ancilla collisions (right) is shown. The addition of these collisions is what introduces non-Markovianity into the dynamics. The basic collision model gives rise to quantum homogenisation.
}
\label{collisions}
\end{figure}

Our model consists of sequential system-environment interactions interspersed with subenvironment-subenvironment couplings, according to the general scheme illustrated in Fig.~\ref{collisions}: after the evolution induced on the joint state of $S$ and the $j^\text{th}$ subenvironment by $\hat{\cal S}_{S,j}$, the system shifts by one site in the lattice while the $j^\text{th}$ environmental two-level system interacts with its rightmost nearest neighbour. Given the generic factorized initial state $\rho^{SE}_0$%=\rho^S_0\otimes\rho^E_{0}$ with $\rho^{S(E)}_0$ the initial state of the system (super-environment)
, the overall evolution after interaction with $n$ elements of the super-environment can be formally written in terms of the unitary map %|{\{0\}}\rangle\langle{\{0\}}|_{\{E\}}$ 
%
%We thus have at the nth iteration the system-bath composite
\begin{equation}
\rho^{SE}_n\equiv\hat U_n\rho^{SE}_0\hat U^\dag_n%\Upsilon[\sigma^{SE}_0]=\hat\Psi_{n,n+1}[\hat\Phi_{S,n}[...\hat\Psi_{1,2}[\hat\Phi_{S,1}[\sigma^{SE}_0]]...]]
\end{equation}
with %$\hat\Upsilon[\cdot]$ 
$\hat U_n$ the overall unitary evolution experienced by the $S$-$E$ system that is generated by the composition of the set of unitary gates introduced above.
The dynamics of $S$ is then retrieved by discarding any information on the degrees of freedom of the super-environment, which breaks down the unitarity of the overall process. However, as we will discuss later on in this paper, the way such a process is accounted for is actually quite crucial and represents a very subtle point to consider. % $\rho^S_n=\text{Tr}_E[\sigma^{SE}_n]$.

To start making quantitative statements on the problem at hand, it is convenient to introduce one of the key instruments that will be used throughout this paper, namely the measure for non-Markovianity proposed in Ref.~\cite{Breuer09}. This is based on the study of the time-behavior of the trace distance between two generic states,
\begin{equation}
\label{tracedistance}
D(\rho_1,\rho_2)=\frac12\|\rho_1-\rho_2\|_1,
\end{equation}
with $\|\cdot\|_1$ the trace norm~\cite{Nielsen}. The trace distance is equal to $1$ for fully distinguishable states and is null for identical states. If we take $\rho_{1,2}$ as the states of a system that is evolving under the action of a dynamical map starting from two different initial states, the degree of non-Markovianity is defined as 
\begin{equation}
\label{measure}
{\cal N}=\max_{\{\rho_1(0),\rho_2(0)\}}\int_{\Omega_+}\partial_t D(\rho_1(t),\rho_2(t)) dt
\end{equation}
where $\Omega_+=\bigcup_i(a_i,b_i)$ is the union of all the time intervals $(a_i,b_i)$ in our observation window within which $\partial_t D(\rho_1(t),\rho_2(t)) >0$, $\rho_{1,2}(0)$ are two initial states of the system and $\rho_{1,2}(t)$ is their time-evolved form. The maximisation is performed over all possible pairs of initial states. The function $\partial_t D(\rho_1(t),\rho_2(t))$ encompasses the condition for revealing the non-Markovianity of an evolution: the existence of even a single region where $\partial_t D(\rho_1(t),\rho_2(t))>0$ is sufficient to guarantee non-Markovian nature of a dynamics. Conceptually, in fact, ${\cal N}$ accounts for all the temporal regions where the distance between two arbitrary input states increases, thus witnessing a re-flux of information from the environment to the system under scrutiny. Such a re-flux magnifies the difference between two arbitrarily picked input states evolved up to the same instant of time by the same map. Contractivity of the trace distance under divisible maps ensures that such re-flux never occurs, which in turn is equivalent to $\partial_t D(\rho_1(t),\rho_2(t))\le0$. As the evolution under scrutiny here proceeds in discrete temporal steps, we will employ the discretised version of Eq.~(\ref{measure}), which is obtained as
\begin{equation}
{\cal N}=\max\sum_n[D(\rho^S_{1,n},\rho^S_{2,n})-D(\rho^S_{2,n-1},\rho^S_{2,n-1})]
\end{equation} 
with $\rho^S_{k,n}$ the state of system $S$ obtained starting from the initial state $\rho^S_{k,0}$ after $n$ steps of our protocol. In the following, we will use the system preparation %short-cut notation $\rho^S_{k,0}\to\rho^S_{\theta_k}$ with 
\begin{equation}
\rho^S_{k}=
\begin{pmatrix}
\cos^2\theta_k&\cos\theta_k\sin\theta_k\\
\cos\theta_k\sin\theta_k&\sin^2\theta_k
\end{pmatrix},~~~\theta_k\in[0,2\pi]
\end{equation} 
standing for a pure initial state of $S$ determined by the angle $\theta_k$ in the Bloch sphere. 

\subsection{Quantum homogeneization}
Armed with such a tool, we start noticing that, without introducing inter-ancilla collisions in our process, i.e. for $\delta=0$, the model describes the process of {\it quantum homogenisation}~\cite{Buzek01} whereupon preparation of an identical fiducial state for all the elements of the super-environment, the state of the system eventually homogenises to it, thus realizing a microscopic model for Markovian decoherence.
\begin{figure}[t]
{\bf (a)}\hskip4cm{\bf (b)}
\includegraphics[width=\columnwidth]{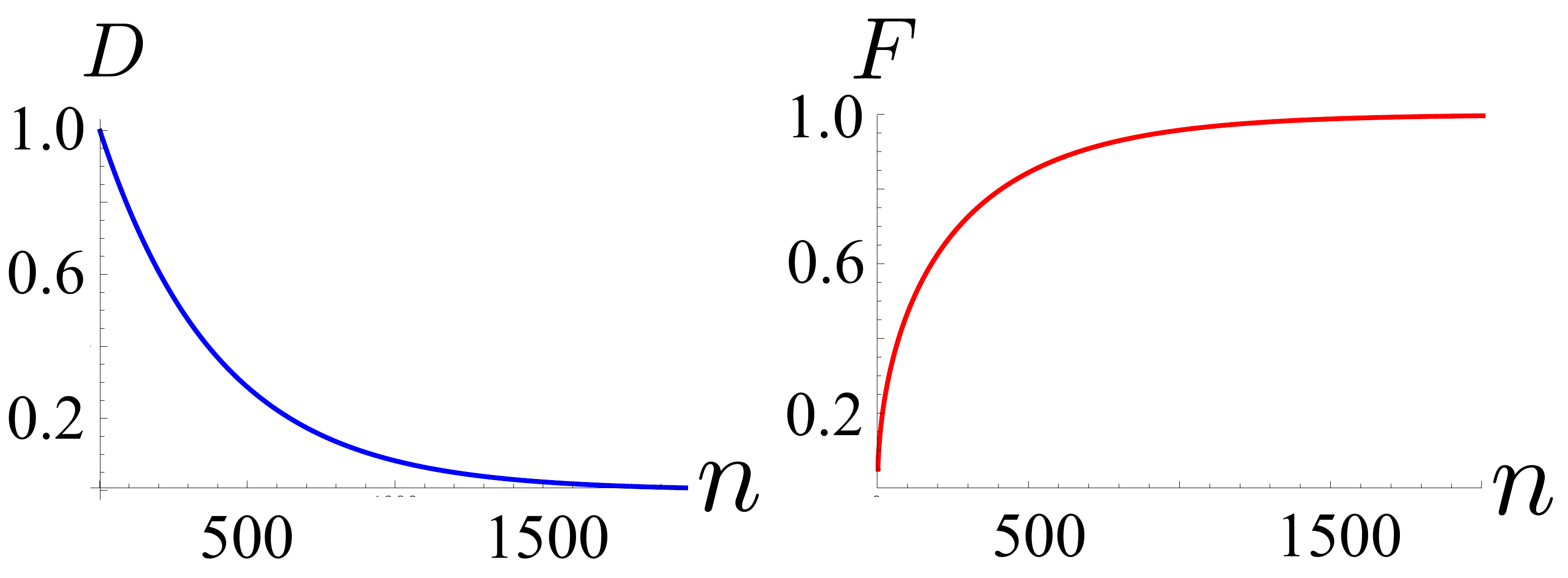}
\caption{The trace distance between evolved system states [panel {\bf (a)}] and the fidelity of the system state at step $n$ with the target one [panel {\bf (b)}] are shown against the number of steps of the evolution for the case of the basic collision model that forbids inter-ancilla interactions and a super-environment prepared in $\ket{\{0\}}$. We have taken the swap strength $\gamma=0.05$, while the initial states used to calculate the trace distance in panel {\bf (a)} correspond to $\theta_1=\pi/2$ and $\theta_2=0$. }
\label{homogeneous}
\end{figure}
In order to set a useful benchmark for the comparison with quantum homogenisation, in our assessment of the non-Markovian features arising from the inclusion of intra-environment  interactions, we will also consider an identical fiducial state for the super-environment. In particular, we will consider the initialisation $\ket{\{0\}}\equiv\otimes^N_{j=1}|0\rangle_{j}$. In Fig.~\ref{homogeneous}, we show the behaviour of trace distance $D$ and state fidelity $F={}_S\langle0|\rho^S_{n}|0\rangle_S$ between the state of the system after $n$ collisions with the super-environment and the target state $\ket{0}_S$ (i.e. the state into which each element of the environment is prepared). The trace decreases monotonically with the number of system-environment collisions, thus witnessing the complete absence of any back-flow mechanisms that might give rise to non-Markovian features. These conclusions hold qualitatively regardless of the initial preparation of the super-environment.

%where $\sigma_0=\rho_0|{\bf0}\rangle\langle{\bf0}|_B$ is the system-bath initial state with $\rho_0$ the density matrix of the input system state and $|{\bf0}\rangle_B=|0\rangle_1|0\rangle_2...|0\rangle_n$.

\subsection{Description of the strategies}

We are now in a position to attack the main goal of this paper and address the delicate point of tracking the reduced dynamics of the system $S$. We will focus on two inequivalent ways of tracing out the degrees of freedom of the super-environment. In turn, this will allow us to pinpoint the % To compute this we will proceed with two different iterative methods in order to demonstrate the 
key role played by SECs in the settling of non-Markovian features. 

The first method that we use in order to compute the reduced dynamics of $S$ (which we dub {\it \sone}) is that of tracing out one of the subenvironments as soon as the system has interacted with it. This corresponds to taking a ``utilitarian" viewpoint according to which element $E_j$ of the super-environment is relevant to determining the evolution of $S$ only as far as their mutual interaction is concerned. This means that correlations between the $S$ and $E_j$ are erased before the interaction between $E_j$ and $E_{j+1}$ occurs and cannot, therefore, affect $S$ during the next iteration of interactions. In this respect, the effect that the ``collision" with $S$ has on the state of $E_j$ is carried over across the super-environment regardless of the actual state of $S$ itself. That is, the reduced state of the system qubit after the interaction with a set of $n$ subenvironments is described by the dynamical map 
\begin{equation}
\label{one}
\rho^S_n=\text{Tr}_{n-1,n}(\hat{\Psi}_{n-1,n}[\hat{\Phi}_{S,n}[\rho^S_{n-1}\otimes \text{Tr}_{S}(\rho^{SE}_{n-1})\otimes \ket{0}\!\bra{0}_n]]),
\end{equation}
with $\text{Tr}_{S}(\rho^{SE}_{n-1})$ the reduced state of the $(n-1)^\text{th}$ element of the subenvironment after its interaction with the system qubit, which is in turn left in state $\rho^S_{n-1}$.

The second approach (named {\it \stwo} hereafter) implies the tracing out  of a subenvironment only after it has outlived its usefulness. In more explicit terms, we consider the various subenvironments in a time-non-local fashion: element $E_j$ will be traced out only after its active role in the collision model has expired, i.e. when it has interacted with the ordered triplet ($E_{j-1}, S, E_{j+1}$). The dynamical map resulting from the implementation of this strategy thus gives rise to the reduced state of the system,
\begin{equation}
\label{two}
\rho^S_n=%\Upsilon[\rho^S]\equiv\text{Tr}_n[\sigma_n]=
\text{Tr}_{n-1,n}(\hat{\Psi}_{n-1,n}[\hat{\Phi}_{S,n}[\text{Tr}_{\{n-2\}}(\rho^{SE}_{n-1})\otimes\ket{0}\!\bra{0}_n]]),
\end{equation}
where $\text{Tr}_{\{n-2\}}[\cdot]$ denotes the partial trace over the whole set of $n-2$ elements of the super-environment prior to the interaction between $S$ and element $E_{n-1}$. %$\varsigma^S_n$ is the two qubit system-ancilla ensemble at the nth iteration, Tr$_{n}[\sigma_n]$ is the partial trace of the environment qubit that will no longer play any part in the dynamics leaving the reduced density matrix for the remaining two qubit composition between the system qubit and the environment qubit it has yet to directly interact with, and $e_0$ denotes the environment ancilla prepared in the $|0\rangle$ state.The second scheme deals with the density matrices of the system and environment qubits separately and does not preserve the SECs between them. It uses partial traces to separate all the qubits erasing all correlations: where Tr$_{n,n+1}[\sigma_n]$ is the partial trace of the two environment qubits leaving only the reduced density matrix for the remaining system qubit, $\rho_n$, Tr$_{\rho,n}[\sigma_{n-1}]$ is the partial trace of the system qubit and the second environment qubit leaving only the reduced density matrix of the first environment qubit, and $e_0$ denotes the environment ancilla prepared in the $|0\rangle$ state.
Quite intuitively, the difference between the two strategies resides in the different way SECs are treated: while $\sone$ erases all the SECs established by a given system-subenvironment collision, the second one carries these over to the next intra-environment interaction. This results in considerable differences in the non-Markovianity features arising from the  dynamical maps  formalised by Eq.~\eqref{one} and \eqref{two}. However, contrary to a naive expectation, $\sone$ does not give rise to a homogenisation process such as the one addressed earlier on, in light of a non negligible environmental memory effect, and it leaves room for non-Markovian manifestations. We have checked that the qualitative features that will be showcased throughout our analysis depend critically on the degree of purity of the overall super-environmental state, but only weakly on its explicit form. %{\bf MAURO: Ruari, can we comment a bit on this? What happens if we change the state of the environment? Surely, if we have a fully mixed state, markovianity is re-instated. But how about taking a differrent pure state?}%.
%We obtain maximally non-Markovian behaviors when the system and super-environmental states . %If we take for the bath an ensemble of fully mixed states then we will see Markovian dynamics. If, instead, we take a different pure state for the bath we will observe non-Markovian dynamics similar to what will be presented later, but the non-Markovianity will be capped by the initial distinguishability between the system and environment states. This is why in our treatment we use orthogonal states.

\section{Analysis of the degree of non-Markovianity}
\label{number}

Here we present our analysis of the non-Markovian features of the collision model, addressing both of the strategies identified above. 

\subsection{Non-Markovianity resulting from both Strategies}

\label{nonmd}

\begin{figure}[!t]
%{\bf (a)}
%\includegraphics[width=\columnwidth]{Strategy2}\\
%{\bf (b)}
%\includegraphics[width=\columnwidth]{Strategy1}
{\bf (a)}\hskip4cm{\bf (b)}\\
\includegraphics[width=\columnwidth]{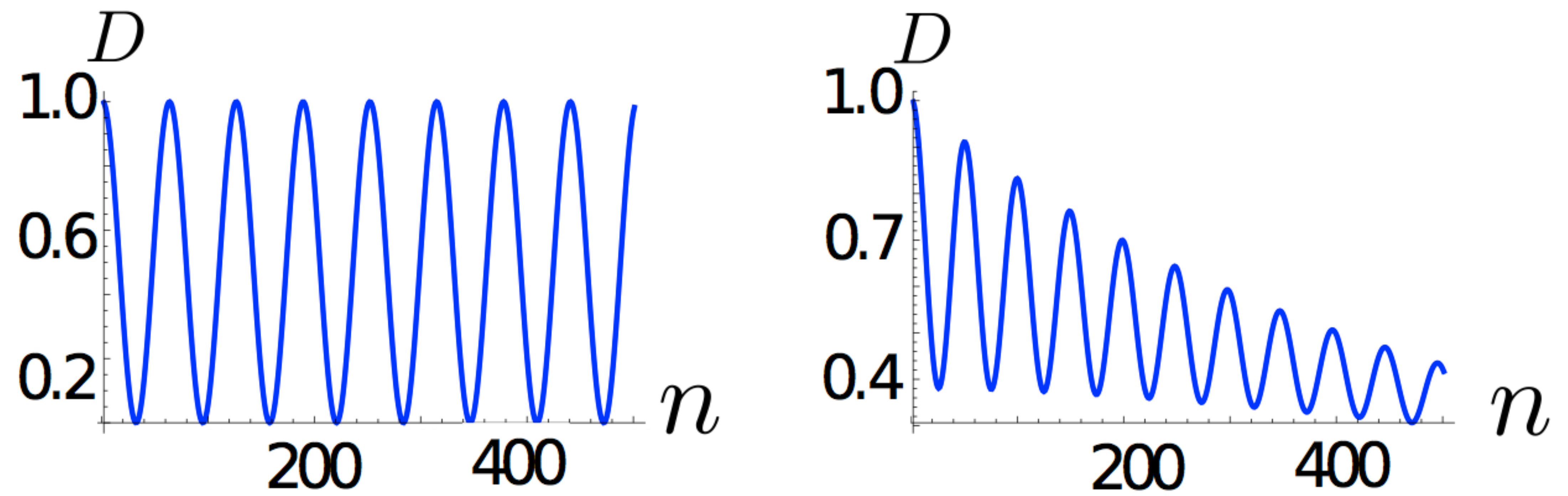}\\
{\bf (c)}\hskip4cm{\bf (d)}\\
\includegraphics[width=\columnwidth]{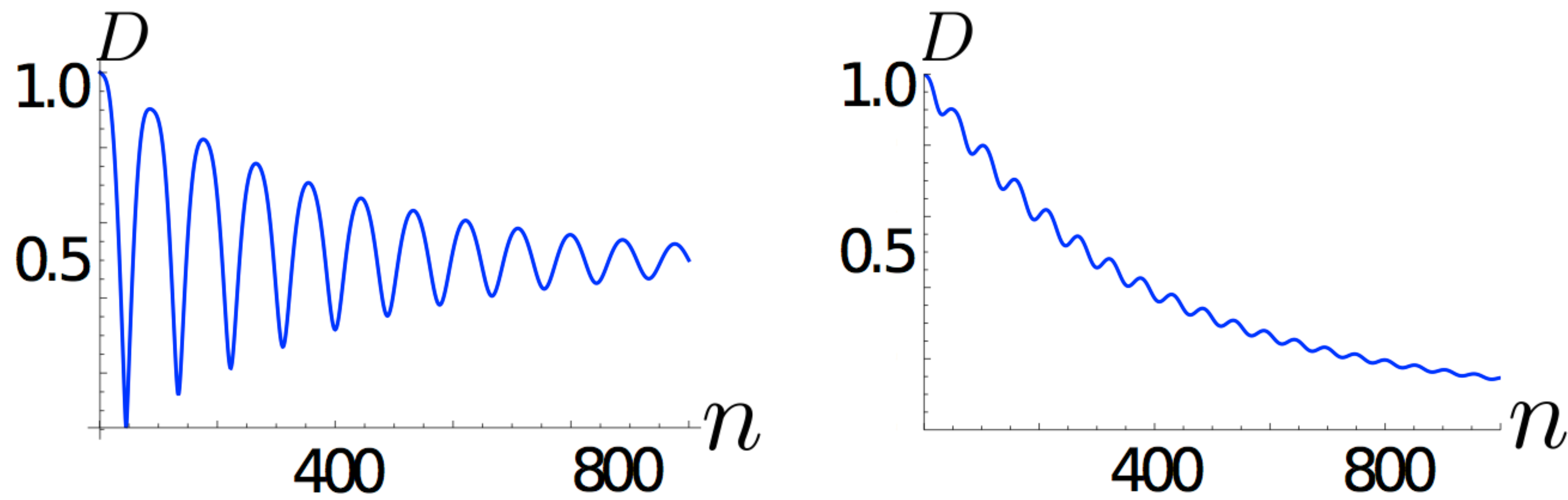}
\caption{%{\bf MAURO: suggestions: let's reduce the number of steps for both the plots. 500 steps for the upper plot, 10000 for the lower one. you use the 30000-step one as an inset for panel (b). In panel (b), put the horizontal axis at D=0.}%
The trace distance $D$ plotted against the number of iterations of the collision model for both the strategies and for different swap strengths put in place in our analysis. Panels {\bf (a)} and {\bf (b)} [{\bf (c)} and {\bf (d)}] report the results valid for \stwo [\sone]. %The pair of angles $(\theta_1,\theta_2)$ that is needed in order to calculate the measure of non-Markovianity employed in this paper are those giving the $|0\rangle$ and $|1\rangle$ states for Panel {\bf{(a)}} and {\bf{(b)}} and $|0\rangle$ and $|0.895\rangle$ for Panel {\bf{(c)}} and {\bf{(d)}}. 
We have used $\delta={\pi}/{2}$ in panels {\bf{(a)}} and {\bf{(c)}} and $\delta=0.95 \times {\pi}/{2}$ in panels {\bf{(b)}} and {\bf{(d)}} to show how varying the inter-ancilla swap strength changes substantially the degree of non-Markovianity. We have taken $\gamma=$0.05 for the system-environment interaction strength, consistently with the assumption of weak $S$-$E$ coupling. }
\label{nonmmeas1}
\end{figure}

We start analysing the behaviour of the trace distance $D(\rho_{\theta_1},\rho_{\theta_2})$ as the collision-based model for system-environment interaction is iterated. %{\bf MAURO: Ruari, here we need a little more study on the behavior of the trace distance. Can we show what happens for a varying $\delta$, please?} %
When using $\stwo$ with $\delta={\pi}/{2}$ (i.e. for a full state-swap between two consecutive super-environmental elements), the joint dynamics of $S$ and $E$ resembles that of an iterated two-qubit system. % {\bf MAURO: this is an important point, but we need to address it more extensively. We should explain better and show what happens if we change $\delta$}.
 This is due to the complete exchange of information between environment qubits at every step, which causes the system to interact at step $n$ with a {\it fresh} physical information carriers that, however, carries fully the effect of the collision occurred at step $n-1$. This results, needless to say, in dynamics characterised by undamped oscillations of the trace distance, which would in turn give rise to a degree of non-Markovianity that would grow with the size of the temporal window of observation of the evolution reaching, asymptotically, an infinite value [cf. Fig.~\ref{nonmmeas1}{\bf (a)}]. 
 
 However, requiring $\delta=\pi/2$ implies, in quite a general sense, a strong intra-environment interaction. Unsurprisingly, this results in the non-Markovian features highlighted above, in light of the pronounced dynamical nature of the corresponding super-environment. It is thus interesting to address the case of $\delta<{\pi}/{2}$, i.e. a weaker subenvironment-subenvironment coupling strength. The expectation is that this would correspond to a loss of information over the state of the $n^{\text{th}}$ environmental element, whose state is only partially carried over to element $(n+1)^{\text{th}}$.  This is well captured by the trace distance, which oscillates with a degraded amplitude, as illustrated in Fig.~\ref{nonmmeas1}{\bf(b)}. %In Fig~\ref{nonmmeas1}{\bf(a)} the constant amplitude of the peaks shows the accuracy of the two qubit system-environment ensemble analogy and in Fig~\ref{nonmmeas1}{\bf(b)} the amplitude damping shows the information loss from the incomplete swaps between environment qubits. %{\bf MAURO: this should go in the part where the measure is described We may integrate the area under this graph, in the periods where the trace distance derivative is positive, to find the actual numerical measure of non-Markovianity by Breuer {\it et al.}~\cite{Breuer09}}.

\begin{figure}[t]
\includegraphics[width=0.8\columnwidth]{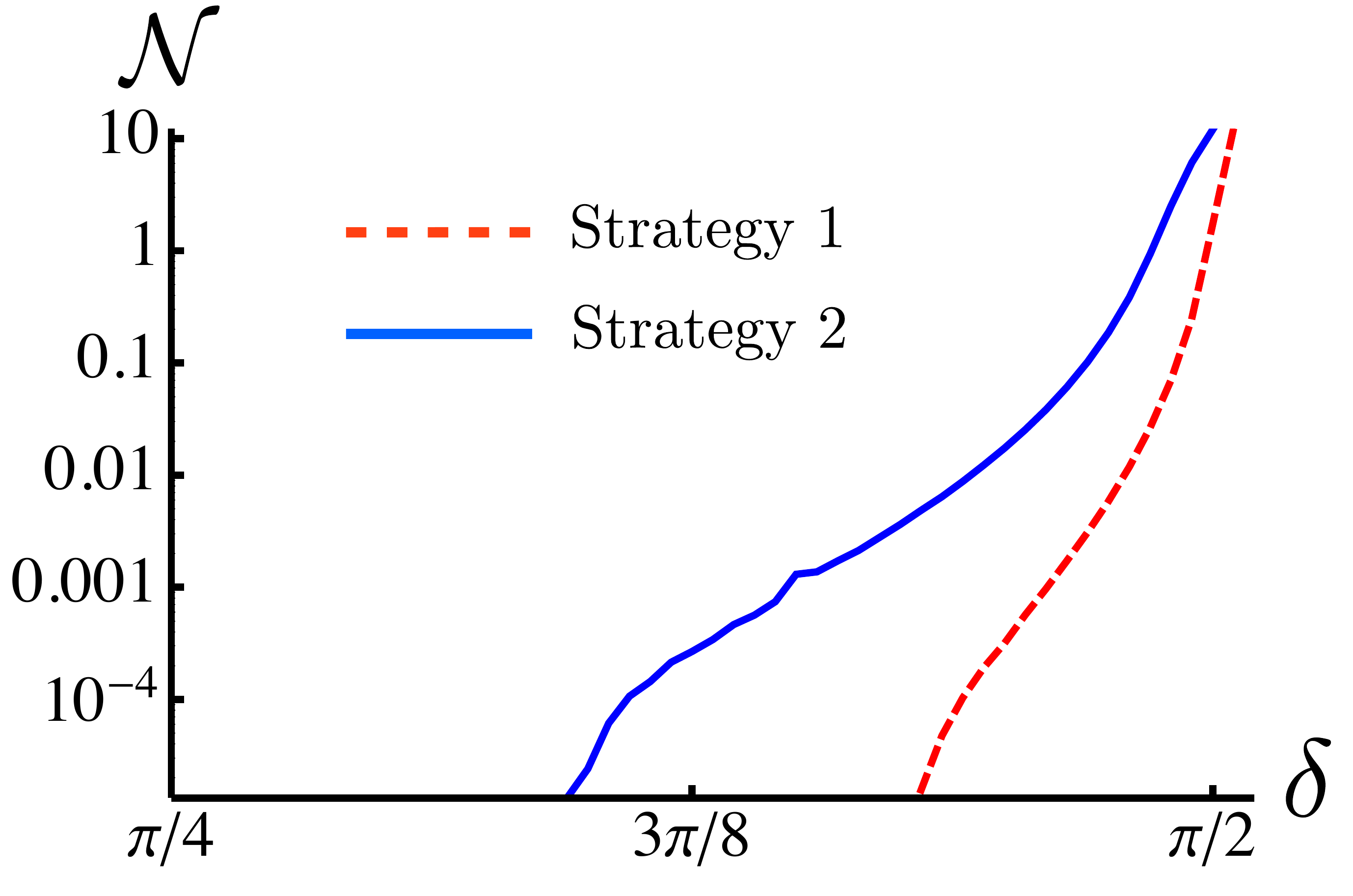}
\caption{Measure ${\cal N}$ plotted against the intra-environment interaction strength $\delta$, for both the strategies addressed in this work. We have used $\gamma=0.05, n=3\times10^4$, and  the super-environment is initialised in $\ket{\{0\}}$. By approaching the full-swap condition embodied by $\delta={\pi}/{2}$, the non-Markovianity measure shoots up. The vertical axis of the plot has been truncated to ${\cal N}=10$ so as to improve the visibility of the details of the plot. The qualitative and quantitative differences inherent in the different strategies for the modelling of $S$-$E$ interactions are evident in the different thresholds in the value of $\delta$ above which ${\cal N}>0$. The measure is optimised over all possible pairs of initial $S$ states.}  %with respect to the change of inter-ancilla collision strength. The initial states entered are the orthogonal states for the SEC model, and the computationally optimised states for the SEC-cancelling model. }
\label{logplot}
\end{figure}

The trend shown in Figs.~\ref{nonmmeas1}{\bf (c)} and {\bf (d)} highlights the differences between the two strategies addressed in our study. In fact, when \sone is employed to model the $S$-$E$ interaction, even the strongest  intra-environmental coupling strength produces a depleted back-flow mechanism:  %this figure the  alternative scheme in which SECs are cancelled is used. In the complete swap case, discussed for the first scheme above, we expect to see periodic, undamped oscillations in the trace distance forever with no loss of information, however, this is not what we see in the case where SECs are cancelled at every stage. First of all, there is zero non-Markovianity in the case of orthogonal states, so in Fig~\ref{nonmmeas1}a we have presented the trace distance against the interactions for the input state which optimises the non-Markovianity of the system. Secondly, 
the initially significant oscillations of the trace distance gradually fade to a small yet non-null value. While the dynamics persists to be non-Markovian even under \sone, the qualitative features of the evolution are indeed strongly dependent on the way information is propagated across.  As we will argue in the following Section, the differences between the two dynamical strategies are due to the different way SECs are accounted for. %{\bf MAURO: Ruari, the following sentence is not very clear Interestingly, for $\delta<\pi/2$ the information loss eventually leads to a homogenisation to the bath state, in our case $|{\bf 0}\rangle$, but the information loss due to the SEC cancelling leads instead to tiny oscillations around the maximally mixed state}.
Also, while for the case of  {\it strategy 2} the maximum inherent in the definition of the measure of non-Markovianity used here is achieved for the input states $|0\rangle$ and $|1\rangle$, for {\it strategy 1} we require the pair $({|0\rangle-|1\rangle})/{\sqrt{2}}, ({|0\rangle+|1\rangle})/{\sqrt{2}}$. We can quantify the difference between the two schemes by putting in place the measure stated in Eq.~\eqref{measure} for different values of $\delta$. In Fig.~\ref{logplot} the results associated with the two schemes are shown for 100 different values of $\delta\in[0,{\pi}/{2}]$. Strategy 2 is spectacularly superior to \sone in setting a non-zero degree of non-Markovianity even at moderate values of the $S$-$E$ interaction, and it has a comparably smaller threshold in the value of $\delta$ above which the dynamics is signalled as non-Markovian. For $\delta=\gamma={\pi}/{2}$, ${\cal N}=n$. In this case, in fact, we have complete swaps at every iteration in both system-environment and intra-environment interactions. 

%Since the SECs maintaining model displays more overall non-Markovianity the rest of this paper will deal largely with this model alone as it is potentially more useful and interesting in terms of SECs. However there will be some more discussion about the differences between the models investigated with the methods presented in the next section. 

\subsection{The role of SECs}\label{SEC}

So far, the importance of SECs in establishing non-Markovian features in our model has been only hinted at without a rigorous quantitative justification. We now fill this gap by using a recently proposed framework that can provide an upper bound to the changes of the trace distance based on the amount of SECs in the state at hand~\cite{Mazzola12,Smirne}. By calling $\beta(t,\rho^S_{1,2})=\partial_t D(\rho^S_1,\rho^S_2)$ and dropping the iteration label for ease of notation, such upper bound is formally given by
%\begin{widetext}
\begin{equation}\label{upper}
\begin{aligned}
\beta\left(t,\rho_{1,2}^S\right)&\leq\frac{1}{2}\left( \min\limits_{k=1,2} \left\|\text{Tr}_E\left[\hat{H},\rho_k^S(t)\otimes(\rho_1^E(t)-\rho_2^E(t))\right]\right\|\right.\\
&\left.+ \left\|\text{Tr}_E\left[\hat{H},\left(\chi_1^{SE}(t)-\chi_2^{SE}(t)\right)\right]\right\|\right).
\end{aligned}
\end{equation}%\end{widetext}
Here $\rho^E_k(t)\equiv\rho^E_{k,n}(t)$ is the reduced state of the environment after $n$ iterations corresponding to the preparation of state $\rho^S_{k,0}$ for the system, and $\chi_j^{SE}(t) = \rho_j^{SE}(t) - \rho_j^S(t) \otimes \rho_j^E(t)$ is the $S$-$E$ correlation matrix. The first term of Eq.~\eqref{upper} contains information about the way the environment evolves when different initial states of the system are inputted. The second term deals with the effects due to non-null SECs. We have examined the respective contributions from  the two terms to explore the origin of the non-Markovianity we observed in Section~\ref{nonmd}. The corresponding results are shown in Fig.~\ref{upperbound}:
\begin{figure}[b]
\includegraphics[width=0.9\columnwidth]{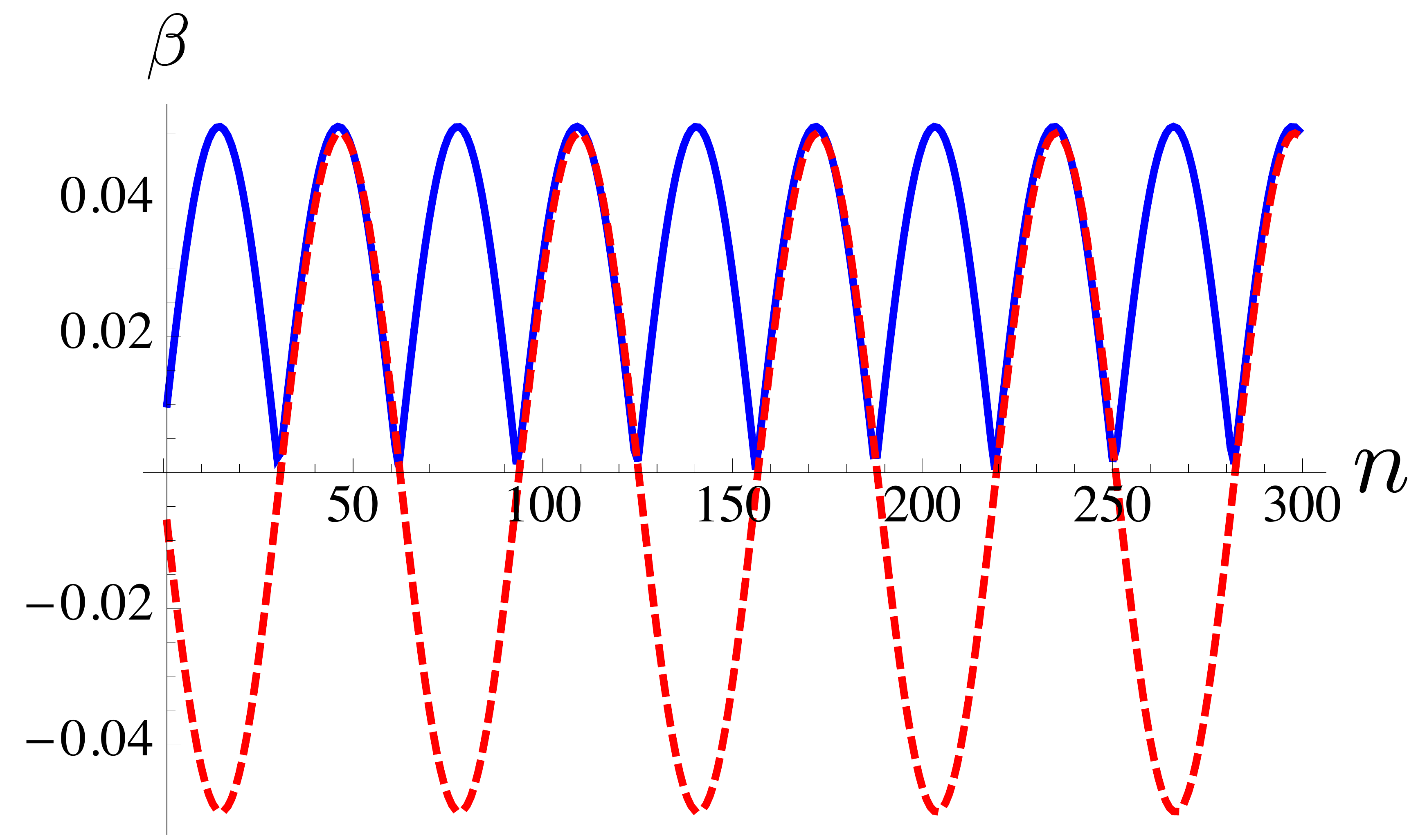}
\caption{The derivative of the trace distance (dashed curve) and the part of the upper bound Eq.~\eqref{upper} dependent on SECs (solid curve) plotted against $n$ for $\delta ={\pi}/{2}$ and $\gamma=0.05$.  }
\label{upperbound}
\end{figure}
The upper bound is found to be completely formed from SECs, the term corresponding to environmental differences being null when the input system states are mutually orthogonal. In turn, this explains why the derivative of the trace distance corresponding to \sone was maximised for values other than the $|0\rangle$ and the $|1\rangle$ state. Indeed, as the origin of non-Markovianity relies entirely on the establishment of SECs, orthogonal input states in the computational basis in \sone, which cancels all of them, would only give rise to ${\cal N}=0$. It is also worth mentioning that when Eq.~\eqref{upper} is computed using \stwo and the input states that are optimal for \sone, we do observe a contribution to the quantitative value of the upper bound to $\beta\left(t,\rho_{1,2}^S\right)$ coming from the dynamical nature of the environment [i.e. the first term in Eq.~\eqref{upper}]. Such a contribution becomes irrelevant for the optimal case of orthogonal input $S$ states. % However, the maximum non-Markovianity comes from the orthogonal case so, although the environmental differences can account for some non-Markovianity in the non-optimal cases, SECs provide a much larger contribution and solely responsible for the non-Markovianity in the optimal case. There is a limit, of course, to how much we can say about the interplay between these two contributions since the measure of non-markovianity is only defined as being valid for the optimal input states.

%\subsection{Investigation with Monte Carlo methods}\label{monte}
%The next section needs reworded%

\begin{figure}[b]
\includegraphics[width=0.9\columnwidth]{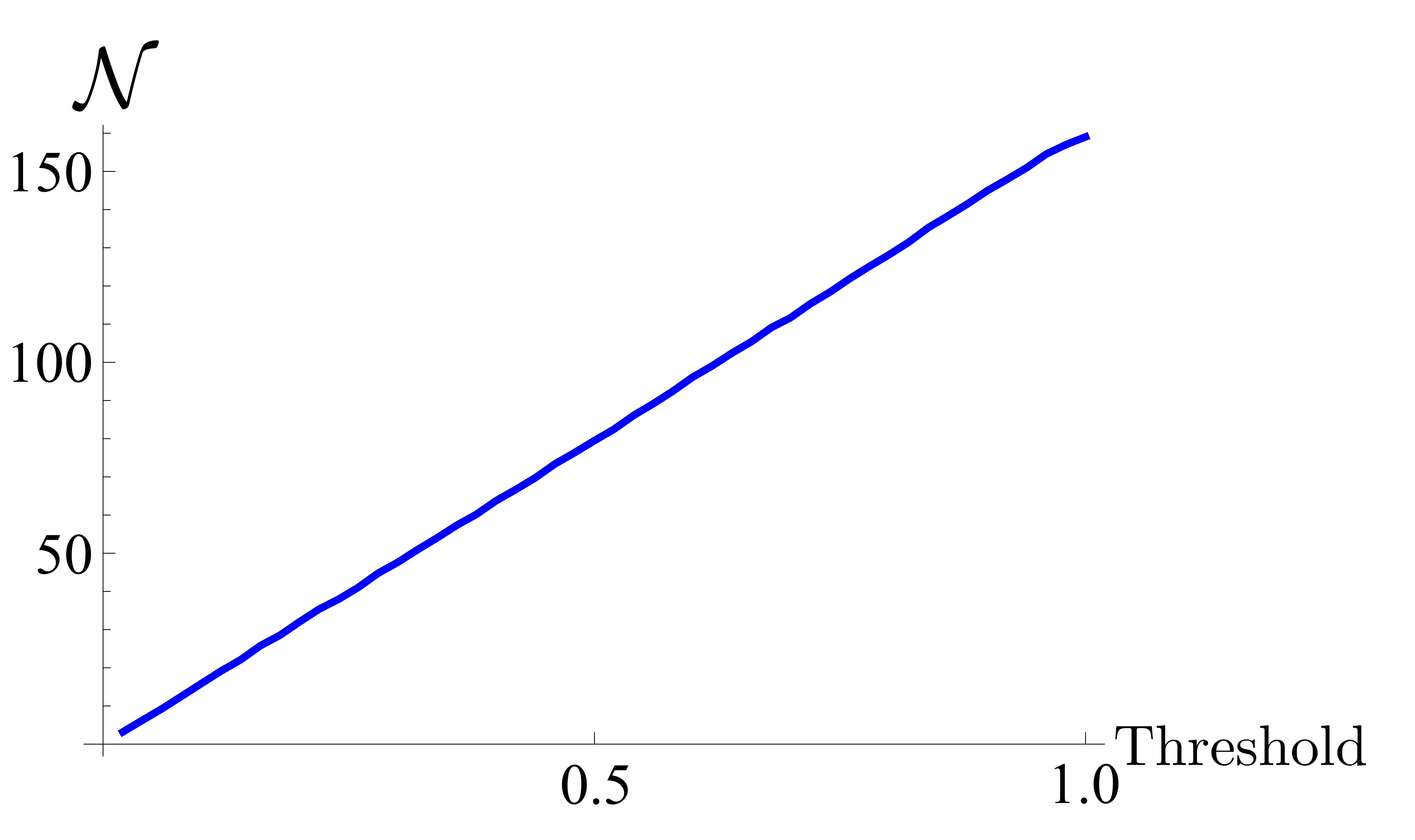}
\caption{The Non-Markovianity measure, $\cal N$ plotted against the strength of the "coin". For this graph we used $\delta=\pi/2$, $\eta=0.01$ and $n=10000$. When we take $\delta=\eta=\pi/2$ we find that ${\cal N}=n$ for the threshold value equal to one. }
\label{montedelta}
\end{figure}

We have extended our numerical analysis by considering aleatory system-environment interactions: in the spirit of Monte Carlo simulations, we have introduced a random variable in our iterative model: should such a variable take a value smaller than a chosen threshold (which we let span the range of values $[0,1]$), $S$ and $E_j$ would interact at step $j$ of the evolution. This process was examined for varying threshold values in Fig.~\ref{montedelta}. Clearly, as the threshold is increased, we allow for system environment interactions to occur, thus increasing the resulting degree of non-Markovianity, which appears to depend linearly on the chosen threshold. 

We have found that, for $\delta={\pi}/{2}$, such a numerical experiment yields changes only in the actual degree of non-Markovianity, which depends on the value taken by the threshold. This is due to the fact that the full swap occurring at the sub-environmental level is not at all affected by a ``missed" system-environment collision: such an event would merely shift the feed-back of the environment into the system to the next ``allowed" interaction. Decreasing the probability of $S$-$E$ interaction will only affect the period of the oscillations of the trace distance, leaving their amplitude unaffected.

\section{Conclusions and outlook}
\label{conc}
We have studied the non-Markovian phenomenology arising from a collision-based microscopic model for system-environment interaction. Our analysis focused on the role that SECs play in the settling of non-Markovian features in the system's dynamics: by putting in place recently proposed tools for the in-depth analysis of the trace distance-based measure of non-Markovianity, and addressing explicitly two non-equivalent iterative protocols for the joint evolution of the system and a multi-particle environment, we have been able to provide evidences of the actual crucial contribution of SECs for the determination of the actual degree of non-Markovianity and the characterisation of the details of such evolution. A Monte Carlo-inspired numerical modelling, built by biasing the chance that a given system-environment interaction actually occurs, showed the persistence of the non-Markovian charactered of the overall evolution. This analysis opens up interesting avenues for the thermodynamic-inspired exploration of non-Markovianity in collision-based models.

\acknowledgments
We thank F. Ciccarello for his helpful insight in the development of this project, Laura Mazzola for her invaluable discussions and Andr\'e Xuereb for his constructive suggestions. RMcC acknowledges financial support from the Northern Ireland DEL. MP thanks the UK EPSRC for support through a Career Acceleration Fellowship and a grant awarded under the ``New Directions for Research Leaders" initiative (EP/G004579/1), the Alexander von Humboldt Stiftung, the John Templeton Foundation (grant 43467), and the EU project TherMiQ (grant agreement 618074).

\end{document}